\documentclass[preprint]{aastex}
\shorttitle{Formation of planetesimals}
\shortauthors{Michikoshi et al.}
\keywords{gravitation, instabilities, methods:$n$-body simulation, planets and satellites:formation}

\begin{document}
\title{$N$-Body Simulation of Planetesimal Formation through Gravitational Instability and Coagulation. II. Accretion Model}

\author{Shugo Michikoshi\altaffilmark{1}, 
Eiichiro Kokubo\altaffilmark{1,2}, and
Shu-ichiro Inutsuka\altaffilmark{3} 
}
\altaffiltext{1}{
Center for Computational Astrophysics, National Astronomical Observatory of Japan, Osawa, Mitaka, Tokyo 181-8588, Japan
}
\altaffiltext{2}{
Division of Theoretical Astronomy, National Astronomical Observatory of Japan, Osawa, Mitaka, Tokyo 181-8588, Japan
}
\altaffiltext{3}{
Department of Physics, Kyoto University, Kyoto 606-8502, Japan
}
\email{michikoshi@cfca.jp, kokubo@th.nao.ac.jp, and inutsuka@tap.scphys.kyoto-u.ac.jp}
\begin{abstract}
The gravitational instability of a dust layer is one of the scenarios
 for planetesimal formation.   
If the density of a dust layer becomes sufficiently high as a result of
 the sedimentation of dust grains toward the midplane of a
 protoplanetary disk, the layer becomes gravitationally unstable and
 spontaneously fragments into planetesimals. 
Using a shearing box method, we performed local $N$-body simulations of
 gravitational instability of a dust layer and subsequent coagulation without
 gas and investigated the basic formation process of planetesimals. 
In this paper, we adopted the accretion model as a collision model.
A gravitationally bound pair of particles is replaced by a single
 particle with the total mass of the pair. 
This accretion model enables us to perform long-term and large-scale
 calculations.
We confirmed that the formation process of planetesimals is the same as
 that in the previous paper with the rubble pile models.
The formation process is divided into three stages: 
the formation of non-axisymmetric structures, the creation of planetesimal seeds, and their collisional growth. 
We investigated the dependence of the planetesimal mass on the
 simulation domain size. 
We found that the mean mass of planetesimals formed in simulations is
 proportional to $L_y^{3/2}$, where $L_y$ is the size of the computational
 domain in the direction of rotation.  
However, the mean mass of planetesimals is independent of $L_x$, where
 $L_x$ is the size of the computational domain in the radial direction
 if $L_x$ is sufficiently large. 
We presented the estimation formula of the planetesimal mass taking into
 account the simulation domain size. 

\end{abstract}

\section{Introduction \label{sec:intro}}
According to the standard theory of planet formation, 
planets form from planetesimals, which are kilometer-sized solid bodies.
However, the process of planetesimal formation is still controversial. 
The inward migration of meter-sized bodies due to gas drag is very rapid;
its timescale is only $10^2$ yr \citep{Adachi1976,Weidenschilling1977}.
The growth of dust to planetesimals due to particle adhering seems difficult in the standard disk model.

The gravitational instability scenario is an alternative scenario for this stage \citep{Safronov1969, Goldreich1973}.
Dust particles settle into the midplane of a protoplanetary disk owing to the gravitational force from the central star if the gas flow is laminar.
As the sedimentation of dust particles proceeds, the density at the midplane exceeds the critical density and the dust layer finally becomes gravitationally unstable.
The gravitationally unstable dust layer rapidly collapses by its self-gravity, forming km-sized planetesimals \citep{Goldreich1973,Coradini1981,Sekiya1983}.
The rapid migration of meter-sized bodies can be avoided because the timescale of the gravitational instability is only on the order of a Kepler period.
However, the gravitational instability scenario has a critical issue.
As the sedimentation of dust grains toward the midplane proceeds, 
the vertical velocity shear increases and gives rise to the Kelvin-Helmholtz instability, which makes the dust layer turbulent.
The turbulence prevents dust particles from settling. 
A lot of studies have been carried out on this issue \citep{Weidenschilling1980, Cuzzi1993, Champney1995, Weidenschilling1995, Sekiya1998, Dobrovolskis1999, Sekiya2000, Sekiya2001, Ishitsu2002, Ishitsu2003, Michikoshi2006}. 
However, problems concerning the gravitational instability still remain unsettled.

Although many studies have been conducted on the condition of gravitational instability or its linear regime, 
there has been little study of the nonlinear regime of gravitational instability.
Here we concentrate on the formation process of planetesimals assuming the gravitational instability.
\cite{Tanga2004} performed $N$-body simulations of a gravitationally unstable particle disk.
In their simulation, the optical depth is very small, and the particle disk is unstable, even initially.
If we consider the formation of planetesimals through gravitational instability at $a_0 \simeq 1\mathrm{AU}$, 
the optical depth is high, and the initial velocity dispersion must be large owing to turbulent flow driven by the Kelvin-Helmholtz instability or magnetorotational instability \citep{Balbus1991}.
In our analysis, we will take a close look at initially stable disks with large optical depth.
\cite{Johansen2007} performed simulations using a code that solves the magnetohydrodynamic equations with a three-dimensional grid 
and includes particles with self-gravity.
They mapped the particle density on the grid and solved the Poisson equation using a fast Fourier transform method.
They included collisions among particles as damping the velocity dispersion of the particles within each grid.
They found that particles collapse in an overdense region in the midplane and the gravitationally bound objects form with masses comparable to dwarf planets.

\cite{Michikoshi2007} performed a set of simulations in the self-gravitating collision-dominated particle disks without gas components (hereafter Paper I).
The point-to-point Newtonian mutual interaction was calculated directly.
They adopted the hard and soft sphere models as collision models, 
and examined parameter dependencies on the size of computational domain, restitution coefficient, optical depth, Hill radius of particles, and the duration of collision. 
In the hard sphere model, penetration between particles is not possible \citep[e.g.,][]{Wisdom1988,Salo1991,Richardson1994,Daisaka1999,Daisaka2001}. The velocity changes in an instant at collisions.
In the soft sphere model, particles can overlap if they are in contact \citep[e.g.,][]{Salo1995}.
They found that the formation process of planetesimals is qualitatively independent of simulation parameters if initial Toomre's $Q$ value, $Q_\mathrm{init}>2$. 
The formation process is divided into three stages: 
the formation of non-axisymmetric wake-like structures \citep{Salo1995}, the creation of aggregates, and the rapid collisional growth of the aggregates. 
The mass of the largest aggregates is larger than the mass predicted by the linear perturbation theory (hereafter linear mass $M_\mathrm{linear}$) \citep{Goldreich1973}.
However \cite{Michikoshi2007} could not find the saturation of the growth of the planetesimals in Paper I. 
The mass of the planetesimal formed in the simulation depends on the size of the computational domain.
Almost all mass in the computational domain is absorbed by a single planetesimal, and thus the mass of the planetesimal is determined by the total mass in the computational domain.

In this paper, we use the alternative model of collisions, `accretion model'.
In physical terms, the accretion model corresponds to the more dissipative model than the hard and soft sphere models.
In the accretion model, when two particles collide, 
if the binding condition is satisfied, the two particles are merged into one particle. 
Efficient energy dissipation occurs when two particles merge into one particle.  
In the soft and hard sphere models, 
since large aggregates consist of a lot of small particles and the number of particles does not decrease, the calculation is time-consuming.
If we are not interested in internal states of planetesimals such as rotation or internal density profile, 
we can treat a large aggregate as one large particle.
We expect that the final mass or spatial distribution of planetesimals can be estimated by the accretion model. 
The accretion model has an advantage from the viewpoint of calculation.
The number of particles decreases as the calculation proceeds; 
thereby this model enables us to perform simulations of larger numbers of particles than that in the previous work.
Taking this advantage, we can examine the large computational domains.

In this study, we neglect the effect of gas for the sake of simplicity.  As
many authors investigated, the effect of gas is very important.  
The solid particles drift radially due to gas drag \citep{Adachi1976,Weidenschilling1977},
gas friction helps gravitational collapse \citep{Ward1976,Youdin2005a,Youdin2005b},
gas turbulence prevents and helps concentration of solids \citep{Weidenschilling1993,Barge1995,Fromang2006,Johansen2006a},
and clumps of solid particles form due to streaming instability \citep{Youdin2005,Johansen2006b,Johansen2007}.
But once gravitational instability happens and large aggregates form, as a first step, we may be able to neglect
gas drag because they are so large that they decouple from gas. 
Therefore our gas-free model may be applicable to the stages of gravitational instability and subsequent collisional growth. 

In Section \ref{sec:method}, we explain the simulation method. Our results are presented in Section \ref{sec:result}.
We compare the accretion model with the soft and hard sphere models and investigate the models of large computational domain.
In Section \ref{sec:summary}, we summarize the results.

\section{Methods of Calculation \label{sec:method}}
The method of calculation is the same as that used in Paper I except for the collision model, 
which is based on the method of simulation of planetary rings \citep[e.g.,][]{Wisdom1988,Salo1991,Daisaka1999,Ohtsuki2000}.

We introduce rotating Cartesian coordinates $(x,y,z)$ called Hill coordinates,
the $x$-axis pointing radially outward, the $y$-axis pointing in the direction of rotation, and the $z$-axis normal to the equatorial plane.
The origin of the coordinates moves on a circular orbit with the semi-major axis $a_0$ and the Keplerian angular velocity $\Omega_0$. 
Assuming $|x_j|, |y_j|, |z_j| \ll a_0$, where $(x_j,y_j,z_j)$ is the position of $j$th particle, 
we can write the equation of motion in non-dimensional forms independent of the mass and the semi-major axis $a_0$, 
if we scale the time by $\Omega_0^{-1}$, the length by Hill radius $h a_0$, 
and the mass by $h^3 M_*$, where $h=(2 m_\mathrm{p}/3M_*)^{1/3}$, $m_\mathrm{p}$ is the characteristic mass of 
particles, and $M_*$ is the mass of the central star \citep[e.g.,][]{Petit1986,Nakazawa1988}:
\begin{equation}
\begin{array}{ccccccc}
\displaystyle{\frac{d^2 \tilde x_i}{d\tilde t^2}} & = &  \displaystyle{2\frac{d \tilde y_i}{d\tilde t}} &+&3 \tilde x_i & + & \displaystyle{\sum_{j=1,j\ne i}^N \frac{\tilde m_j}{\tilde r_{ij}^3}(\tilde x_j- \tilde x_i)}, \\
\displaystyle{\frac{d^2 \tilde y_i}{d\tilde t^2}} & = & \displaystyle{-2\frac{d\tilde x_i}{d\tilde t}} & &      & + & \displaystyle{\sum_{j=1,j\ne i}^N \frac{\tilde m_j}{\tilde r_{ij}^3}(\tilde y_j-\tilde y_i)}, \\
\displaystyle{\frac{d^2 \tilde z_i}{d\tilde t^2}} & = &                   &-&\tilde z_i   & + & \displaystyle{\sum_{j=1,j\ne i}^N \frac{\tilde m_j}{\tilde r_{ij}^3}(\tilde y_j-\tilde y_i)}, \\
\end{array}
\label{eq:EOMx}
\end{equation}
where a tilde denotes corresponding non-dimensional variables, $\tilde r_{ij}$ is the distance between particles $i$ and $j$, and $N$ is the number of particles.

We use the periodic boundary condition \citep{Wisdom1988}.
There are an active region and its copies. 
Inner and outer regions have different angular velocities. 
These regions slide upward and downward with shear velocity $3 \Omega_0 L_x/2$, where $L_x$ is the length of the cell in the $x$-direction.
We calculate gravitational forces from particles in these regions.
We use the special-purpose computer GRAPE-6 for the calculation of self-gravity \citep{Makino2003}.

Particles used in the simulation are not realistic dust particles but superparticles that represent a group of small particles, 
thus the parameter $\epsilon$ used in this Paper is not exactly the physical restitution coefficient 
but corresponds to the coefficient of the dissipation due to collisions among particles.
The collision models used in paper I cannot treat more dissipative models where the tangential coefficients of restitution $\epsilon_\mathrm{t} < 1$
because we adopt $\epsilon_\mathrm{t} = 1$.
The net restitution coefficient is $\epsilon=(\epsilon_\mathrm{n}^2 \langle v_\mathrm{n}^2 \rangle+\epsilon_\mathrm{t}^2 \langle v_\mathrm{t}^2 \rangle)/(\langle v_\mathrm{n}^2 \rangle+\langle v_\mathrm{t}^2 \rangle)$, 
where $\epsilon_\mathrm{n}$ is the normal coefficient of restitution, $v_\mathrm{n}$ and $v_\mathrm{t}$ are normal and tangential velocities of a collision \citep{Canup1995}.
The net restitution coefficient $\epsilon$ is always larger than 0.7 if we assume $\epsilon_\mathrm{t}=1$ and $\langle v_\mathrm{n}^2 \rangle = \langle v_\mathrm{t}^2 \rangle$.
If we allow $\epsilon_\mathrm{t} \ne 1$ in order to examine the model of $\epsilon < 0.7$, we must consider rotations of individual particles.
For the sake of simplicity, we ignore rotations of particles in this paper and assume $\epsilon_\mathrm{t} = 1$.

The way to treat a collision is to use the same method applied to the hard sphere model, but we consider the binding condition of colliding particles.
If a pair overlaps, $\tilde r_{ij} < \tilde r_{\mathrm{p},i} + \tilde r_{\mathrm{p},j}$ where $\tilde r_{\mathrm{p},i}$ is a radius of particle $i$, 
and is approaching, $\tilde \mathbf n_{ij} \cdot \tilde \mathbf v_{ij}<0$ where $\tilde \mathbf v_{ij}$ is the relative velocity and $\tilde \mathbf n_{ij}$ is a unit vector along $\tilde \mathbf r_{ij}=(\tilde x_j, \tilde y_j,\tilde z_j) - (\tilde x_i,\tilde y_i,\tilde z_i)$, 
we assume that this pair collides, and changes its velocities by using the equations of collision:
\begin{equation}
\tilde \mathbf{v_i'}=\tilde \mathbf{v_i}-\frac{\tilde m_j}{\tilde m_i+\tilde m_j}(1+\epsilon)(\tilde \mathbf n_{ij}\cdot \tilde \mathbf v_{ij}) \tilde \mathbf n_{ij},
\end{equation}
\begin{equation}
\tilde \mathbf{v_j'}=\tilde \mathbf{v_j}+\frac{\tilde m_i}{\tilde m_i+\tilde m_j}(1+\epsilon)(\tilde \mathbf n_{ij}\cdot \tilde \mathbf v_{ij}) \tilde \mathbf n_{ij},
\end{equation}
where $\epsilon$ is a restitution coefficient.
Using the updated velocities, we check whether the binding condition is satisfied by the Jacobi integral \citep[e.g.,][]{Nakazawa1988}:
\begin{equation}
  \tilde J_{ij}= \frac{1}{2}(\dot {\tilde x_{ij}}^2+\dot{\tilde y_{ij}}^2+\dot{\tilde z_{ij}}^2)-\frac{3}{2}\tilde x_{ij}^2+\frac{1}{2}\tilde z_{ij}^2-\frac{3}{\tilde r_{ij}}+\frac{9}{2} < 0,
\end{equation}
where $(\dot{\tilde x}_{ij},\dot{\tilde y}_{ij},\dot {\tilde z}_{ij})$ is the relative velocity.
If the pair is bound, two particles are merged into one particle conserving their momentum.
The shape of the new particle is a sphere, the radius of which is determined by its mass.

We assume that all particles have the same mass $m_\mathrm{p}$ initially.
The parameters of the present simulation are the distance from the central star $a_0$, the length of region $L_x$ and $L_y$, the number of particles $N$, and the mass of particles $m_\mathrm{p}(r_\mathrm{p},\rho_\mathrm{p})$, 
where $\rho_\mathrm{p}$ is the density of a particle.
The dynamical behavior is characterized by only two non-dimensional parameters \citep{Daisaka1999}; 
the initial optical depth $\tau$ and the ratio $\zeta=a_0 h/2r_\mathrm{p}$.
In the models of $\zeta>1$, the centers of the particles are in their Hill sphere when two particles come into contact.
The initial optical depth is given by \citep[e.g.,][]{Goldreich1982}
\begin{equation}
\tau=\frac{3\Sigma}{4 \rho_\mathrm{p} r_\mathrm{p}}=0.19 \left(\frac{\Sigma}{10 \mathrm{g\,cm}^{-2}} \right)
\left(\frac{\rho_\mathrm{p}}{2 \mathrm{g\,cm}^{-3}} \right)^{-1}
\left(\frac{r_\mathrm{p}}{20 \mathrm{cm}} \right)^{-1},
\end{equation}
where $\Sigma$ is the surface density of particles. 
The ratio $\zeta$ is given by
\begin{equation}
\zeta=105.528 \left( \frac{M_\odot}{M_* }\right)^{-1/3} \left( \frac{\rho_\mathrm{p}}{2 \mathrm{g}\mathrm{cm}^{-3}}\right)^{1/3}
\left( \frac{a_0}{1 \mathrm{AU}}\right).
\end{equation}

Using the above two parameters, the other parameters can be estimated.
The normalized most unstable wavelength of gravitational instability is \citep[e.g.,][]{Sekiya1983}
\begin{equation}
\tilde \lambda_{\mathrm{most}}=12 \pi \tau \zeta^2.
\end{equation}
The normalized size of computational domain $\tilde L$, the number of particles $N$, the normalized radius of particle $\tilde r_\mathrm{p}$, and Toomre's $Q$ value are given by
\begin{equation}
\tilde L_x = 12 \pi A_x \tau \zeta^2,
\end{equation}
\begin{equation}
\tilde L_y = 12 \pi A_y \tau \zeta^2,
\end{equation}
\begin{equation}
N=576 \pi A_x A_y \tau^3 \zeta^6,
\label{eq:N}
\end{equation}
\begin{equation}
\tilde r_\mathrm{p} = \frac{1}{2 \zeta},
\end{equation}
\begin{equation}
Q= \frac{\tilde \sigma_x}{6 \tau \zeta^2},
\label{eq:qqx}
\end{equation}
where $A_x$ is the ratio $L_x/\lambda_{\mathrm{most}}$, $A_y$ is the ratio $L_y/\lambda_{\mathrm{most}}$, 
and $\tilde \sigma_x$ is the radial velocity dispersion scaled by $a_0 h \Omega_0$, respectively.
In order to determine the size of the computational domain, we need to set $A_x$ and $A_y$. 
The initial velocity dispersion is also a parameter. 
We determine it from the initial Toomre's $Q_\mathrm{init}$ value using Equation (\ref{eq:qqx}).
As shown above, if we set $\tau$, $\zeta$, $A_x$, $A_y$, and $Q_\mathrm{init}$, 
we can determine the initial state.
The simulation parameters are summarized in Table $\ref{table:model2}$.
We call model 1 as the standard model.
The number of particles is 1000-10000.

If $\zeta>1.5$, two particles can be bound \citep{Ohtsuki1993,Salo1995}, so we
need to adopt $\zeta>1.5$ in order to investigate the formation process of planetesimals.
In this paper, we use $\zeta=2$, which is much smaller than the realistic value $\zeta \simeq 100$.  
Thus the physical size of a particle in our simulation is very large.  It indicates
that the effect of the self-gravity is relatively weaker than that of inelastic
collisions.  To investigate gravitational instability, the particle density should be higher than the Roche
density.  The particle density is $\rho_\mathrm{p} = 4 \zeta^3 \rho_\mathrm{R}$,
where $\rho_\mathrm{R} = (9/4\pi) M_*/a_0^3$ is the Roche density.  Thus when $\zeta = 2$, $\rho_p \gg \rho_\mathrm{R}$.  So we
can study gravitational instability with this parameter. But this approximation may change the result
when comparing between $\zeta=2$ and $\zeta=100$ although gravitational
instability occurs in both models.  In the future work, we will perform the
simulation with a larger $\zeta$ value by using next-generation supercomputers.

In most of the models, we set $Q_\mathrm{init}=3$, which corresponds to $\tilde
\sigma_x \simeq 7.2$ for the standard model.  The characteristic velocity of the actual turbulence
$v_\mathrm{turb}$ is about $\eta v_\mathrm{K}$, where $v_\mathrm{K}$ is the Kepler velocity,
$\eta=-12(c_s^2/v_\mathrm{K}^2)(\partial \log P / \partial \log r)$, $c_s$ is the sound
velocity of gas, $P$ is the pressure of gas, and $r$ is the distance from the
central star \citep{Adachi1976,Weidenschilling1977}.  At $r=1\mathrm{AU}$, in the Hill unit, the characteristic velocity of
the turbulence is $\tilde v_\mathrm{turb} \simeq 1.3 \times 10^7$ that is much
larger than $\tilde \sigma_x$.  If the turbulence weakens, $Q$ decreases gradually from
$Q\gg 1$, and gravitational instability occurs finally.  Therefore we choose the
initial condition that the disk is gravitationally stable ($Q_\mathrm{init}=3$).


\begin{deluxetable}{cccccccccc}
  \tablecaption{Initial conditions and Results}
  \tablewidth{0pt}
  \tablehead{
  \colhead{Model} & \colhead{$\epsilon$} & \colhead{$\tau$} & \colhead{$\zeta$}
  & \colhead{$A_x$} & \colhead{$A_y$} & \colhead{$Q_\mathrm{init}$} &
  \colhead{$\sigma_x$} & \colhead{$\langle M_{\mathrm{pl}}/M_{\mathrm{linear}} \rangle$} &
  \colhead{$N_\mathrm{pl}$}
  }
  \startdata 
01   & 0.01       & 0.1   & 2.0    & 6.0 & 6.0 & 3.0 &  7.2 & 21.2 & 2 \\
02   & 0.01       & 0.11  & 2.0    & 6.0 & 6.0 & 3.0 &  7.9 & 42.4 & 1 \\
03   & 0.01       & 0.125 & 2.0    & 6.0 & 6.0 & 3.0 &  9.0 & 21.5 & 2 \\
04   & 0.1        & 0.1   & 2.0    & 6.0 & 6.0 & 3.0 &  7.2 & 42.7 & 1 \\
05   & 0.3        & 0.1   & 2.0    & 6.0 & 6.0 & 3.0 &  7.2 & 43.2 & 1 \\
06   & 0.5        & 0.1   & 2.0    & 6.0 & 6.0 & 3.0 &  7.2 & 38.4 & 1 \\
07   & 0.7        & 0.1   & 2.0    & 6.0 & 6.0 & 3.0 &  7.2 & 0.0 & 0 \\
08   & 0.9        & 0.1   & 2.0    & 6.0 & 6.0 & 3.0 &  7.2 & 0.0 & 0 \\
09   & 0.01       & 0.06  & 2.5    & 6.0 & 6.0 & 3.0 &  6.8 & 43.4 & 1 \\
10  & 0.01       & 0.06  & 2.7    & 6.0 & 6.0 & 3.0 &  8.2 & 43.4 & 1 \\
11  & 0.01       & 0.06  & 3.0    & 6.0 & 6.0 & 3.0 &  9.7 & 39.8 & 1 \\
12  & 0.01       & 0.1   & 2.0    & 4.0 & 4.0 & 3.0 &  7.2 & 19.5 & 1 \\
13  & 0.01       & 0.1   & 2.0    & 8.0 & 8.0 & 3.0 &  7.2 & 38.4 & 2 \\
14  & 0.01       & 0.1   & 2.0    & 3.0 & 3.0 & 3.0 &  7.2 & 11.1 & 1 \\
15  & 0.01       & 0.1   & 2.0    & 6.0 & 3.0 & 3.0 &  7.2 & 21.7 & 1 \\
16  & 0.01       & 0.1   & 2.0    & 9.0 & 3.0 & 3.0 &  7.2 & 16.4 & 2 \\
17  & 0.01       & 0.1   & 2.0    & 12.0 & 3.0 & 3.0 &  7.2 & 11.0 & 4 \\
18  & 0.01       & 0.1   & 2.0    & 24.0 & 3.0 & 3.0 &  7.2 & 10.1 & 9 \\
19  & 0.01       & 0.1   & 2.0    & 3.0 & 6.0 & 3.0 &  7.2 & 21.8 & 1 \\
20  & 0.01       & 0.1   & 2.0    & 3.0 & 9.0 & 3.0 &  7.2 & 32.2 & 1 \\
21  & 0.01       & 0.1   & 2.0    & 3.0 & 12.0 & 3.0 &  7.2 & 43.1 & 1 \\
22  & 0.01       & 0.1   & 2.0    & 3.0 & 24.0 & 3.0 &  7.2 & 85.5 & 1 \\
23  & 0.01       & 0.1   & 2.0    & 6.0 & 6.0 & 4.0 &  9.6 & 43.3 & 1 \\
24  & 0.01       & 0.1   & 2.0    & 6.0 & 6.0 & 5.0 &  12.0 & 21.6 & 2 \\
  \enddata
  \tablecomments{$\langle M_{\mathrm{pl}} \rangle$, $M_{\mathrm{linear}}$ and $N_\mathrm{pl}$ are the mean planetesimal mass, the linear mass and the number of planetesimals at final state, respectively.}
\label{table:model2}
\end{deluxetable}


We set the initial velocities and positions of particles using the above parameters.
A Keplerian orbit has six parameters: the position of the guiding center $(x_\mathrm{g},y_\mathrm{g})$, eccentricity $e$, inclination $i$, and the two phases of epicycle and vertical motions \citep[e.g.,][]{Nakazawa1988}.
The eccentricity $e$ and inclination $i$ of particles are assumed to obey the Rayleigh distribution with the ratio $\sqrt{\langle e^2 \rangle / \langle i^2 \rangle}=2$ \citep{Ida1992}. 
We set the position of the guiding center and the two phases uniformly, avoiding overlapping.

The equation of motion is integrated with a second-order leapfrog scheme.
We adopt the variable time step used by \cite{Daisaka1999}. The time step formula is given by $\Delta t = \eta \min_{i} |\mathbf{a}_i|/|\dot \mathbf{a}_i|$ where $\mathbf{a}_i$ and $\dot \mathbf{a}_i$ are the acceleration of particle
$i$ and its time derivative, and $\eta$ is a non-dimensional time step
coefficient, respectively.

\section{Results \label{sec:result}}

\subsection{Time Evolution}
\begin{figure}
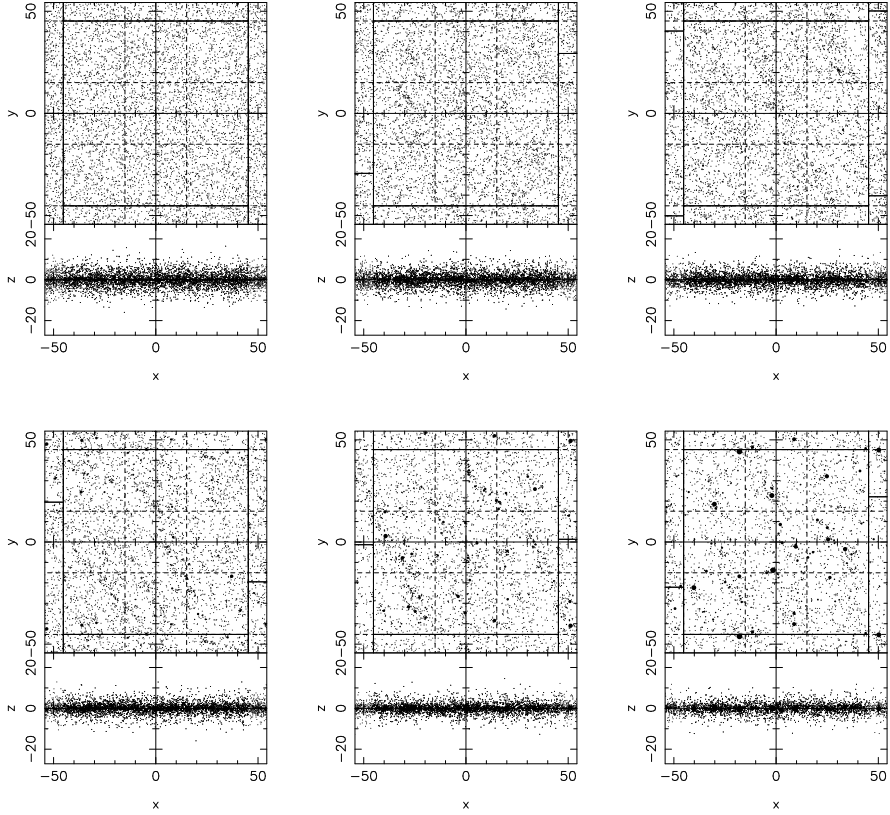

	\begin{center}
		\begin{tabular}{ccc}
			\resizebox{37mm}{!}{\includegraphics{f81.eps}} &
			\resizebox{37mm}{!}{\includegraphics{f82.eps}} &
			\resizebox{37mm}{!}{\includegraphics{f83.eps}} \\
			\resizebox{37mm}{!}{\includegraphics{f84.eps}} &
			\resizebox{37mm}{!}{\includegraphics{f85.eps}} &
			\resizebox{37mm}{!}{\includegraphics{f86.eps}} \\      
		\end{tabular}
		\caption{Spatial distribution of particles in the standard model (model 1) at 
		$t/T_\mathrm{K}=0.0 $ (\textit{top left panel}), 
		$t/T_\mathrm{K}=0.4 $ (\textit{top middle panel}), 
		$t/T_\mathrm{K}=0.8 $ (\textit{top right panel}),
		$t/T_\mathrm{K}=1.2 $ (\textit{bottom left panel}), 
		$t/T_\mathrm{K}=1.6 $ (\textit{bottom middle panel}), and $t/T_\mathrm{K}=2.0$ (\textit{bottom right panel}). Circles represent particles and their size is proportional to the physical size of particles.
		}
		\label{fig:run04pic}
	\end{center}
\end{figure}

Figure \ref{fig:run04pic} shows snapshots in model 1 at $t/T_\mathrm{K}=0.0,
0.4, 0.8, 1.2, 1.6$ and $2.0$ where $T_\mathrm{K}$ is the orbital period $2 \pi
/ \Omega_{0}$.  At $t/T_\mathrm{K}=0.4$, we cannot see any spatial structures
and large bodies.  The formation process of planetesimals is the same as
those of the hard and soft sphere models (Paper I).  The non-axisymmetric
wake-like structure forms at $t/T_\mathrm{K}=0.8, 1.2$, which appears if we
consider self-gravity.  The non-axisymmetric wake-like structure is also seen in planetary rings \citep{Salo1995}.  
The density in the wake-like structure is higher than the mean surface density $\Sigma$. For example, the surface density in the dense region is $1.52 \Sigma$ at $t/T_\mathrm{K} = 1.6$.
Planetesimal seeds form in the dense parts of these wakes by fragmentation ($t/T_\mathrm{K}=1.6, 2.0$).
Here we consider the formation of the wake-like structure and planetesimal seeds as the gravitational instability stage.

\begin{figure}
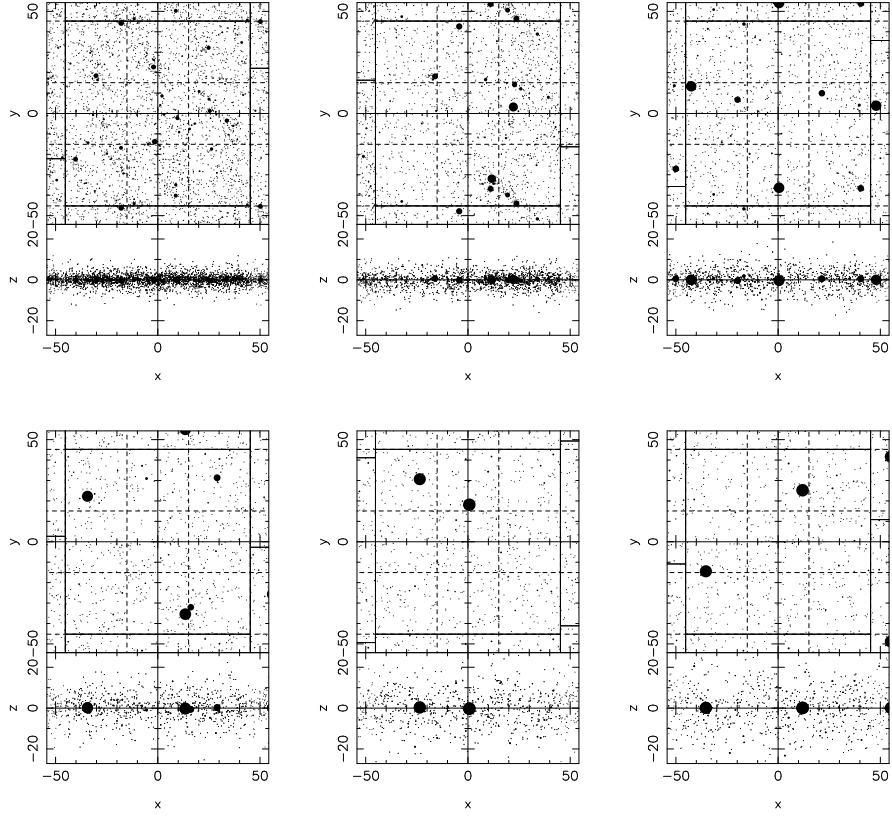

	\begin{center}
		\begin{tabular}{ccc}
			\resizebox{37mm}{!}{\includegraphics{f1.eps}} &
			\resizebox{37mm}{!}{\includegraphics{f2.eps}} &
			\resizebox{37mm}{!}{\includegraphics{f3.eps}} \\
			\resizebox{37mm}{!}{\includegraphics{f4.eps}} &
			\resizebox{37mm}{!}{\includegraphics{f5.eps}} &
			\resizebox{37mm}{!}{\includegraphics{f6.eps}} \\      
		\end{tabular}
		\caption{Same as Figure \ref{fig:run04pic} at
		$t/T_\mathrm{K}=2.0 $ (\textit{top left panel}), 
		$t/T_\mathrm{K}=3.0 $ (\textit{top middle panel}), 
		$t/T_\mathrm{K}=4.0 $ (\textit{top right panel}),
		$t/T_\mathrm{K}=5.0 $ (\textit{bottom left panel}), 
		$t/T_\mathrm{K}=6.0 $ (\textit{bottom middle panel}), and $t/T_\mathrm{K}=7.0$ (\textit{bottom right panel}).
		}
		\label{fig:run04pic2}
	\end{center}
\end{figure}

Figure \ref{fig:run04pic2} shows snapshots in model 1 at $t/T_\mathrm{K}=2.0, 3.0, 4.0, 5.0, 6.0$ and $7.0$.
Once planetesimal seeds form, the non-axisymmetric wake-like structures disappears ($t/T_\mathrm{K}=3.0$).
Planetesimal seeds grow rapidly by mutual coalescence ($t/T_\mathrm{K}=3.0, 4.0, 5.0$).
In this stage, the disk is gravitationally stable. The gravitation seems to merely enhance the coalescent growth.
When almost all planetesimal seeds are absorbed by a few planetesimals, the growth slows down ($t/T_\mathrm{K}=6.0, 7.0$).
This stage is the collisional growth stage.

\begin{figure}
\begin{center}
\includegraphics[width=0.4\linewidth]{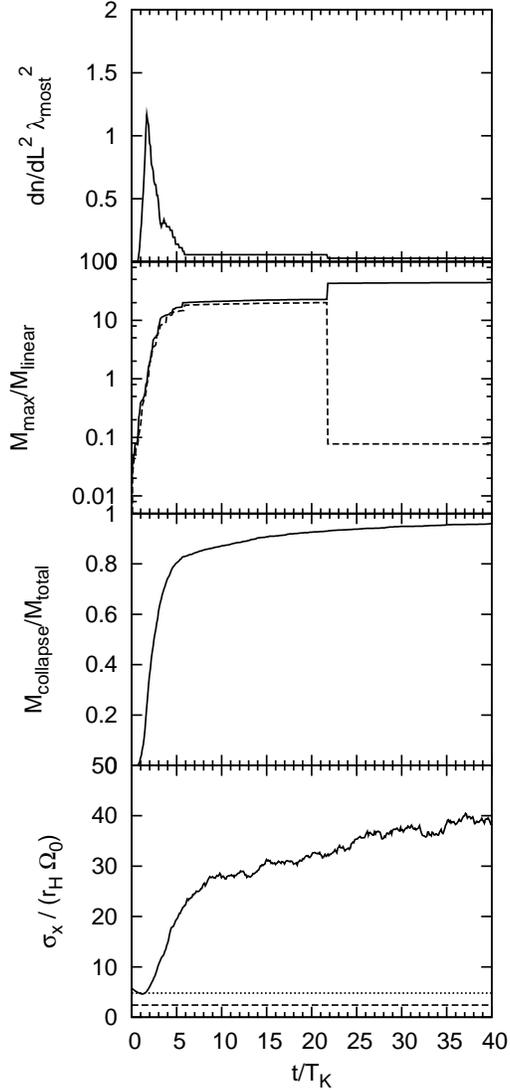} 
\end{center}
\caption{Time evolution of the various quantities in the standard model (model 1).
The quantities are the number of particles with $m> 0.1 M_\mathrm{linear}$ per the area $\lambda _\mathrm{most}^2$, the mass of the largest and second largest bodies $M_\mathrm{max}$ normalized
by the linear mass $M_{\mathrm{lienar}}$, 
the ratio of the total mass of particles with $m > 0.1 M_{\mathrm {linear}}$ to total mass $M_{\mathrm{total}}$ 
and the velocity dispersion of field particles from top to bottom, respectively.
In the bottom panel, we show the velocities for $Q=2$ (\textit{dotted line}) and $Q=1$ (\textit{dashed line}) }
\label{fig:gi2run04a_all.eps}
\end{figure}

Figure \ref{fig:gi2run04a_all.eps} shows the time evolution of the number of
large particles, the mass of the largest and second largest particles, the ratio of the mass of large particles to the total mass and the velocity dispersion of field
particles for model 1.
The time evolution of the number of large particles is similar to those of the hard and soft sphere models (Paper I).  
The number of large particles has a maximum at $t/ T_\mathrm{K} \simeq 1.7$, and its value is about $1.2$, which is slightly larger than that in the hard and soft sphere models.  
However, the number of large particles decreases rapidly after $t/T_\mathrm{K} \simeq 1.7 $.  
This decay is caused by the fast coalescence of large particles. 

The mass of the largest particle monotonically increases up to about $44 M_{\mathrm{linear}}$, 
which is twice larger than that of the hard and soft sphere models, where $M_{\mathrm{linear}} = \pi \Sigma (\lambda_{\mathrm{most}}/2)^2$ 
is the planetesimal mass estimated by the linear theory.
In the accretion model, the growth of planetesimals is enhanced because sticking of dust grains is efficient.  
The mass of the second largest particle changes discontinuously, 
when the second largest particle is absorbed by the largest one.

The ratio of the mass of the large particles to the total mass is almost the same as that of the hard and soft sphere models.  
The ratio monotonically increases up to about 1,
and most of the mass in the computational domain is finally absorbed by the large planetesimals.

The velocity dispersion decreases until $t/T_\mathrm{K} \simeq 1.3$ because of the dissipation due to inelastic collisions. 
When the velocity dispersion becomes sufficiently small ($Q\simeq 2$), the gravitational instability occurs and
a lot of planetesimal seeds form.
As the field particles are scattered by the planetesimal seeds, 
the velocity dispersion increases.

\begin{figure}
\begin{center}
\includegraphics{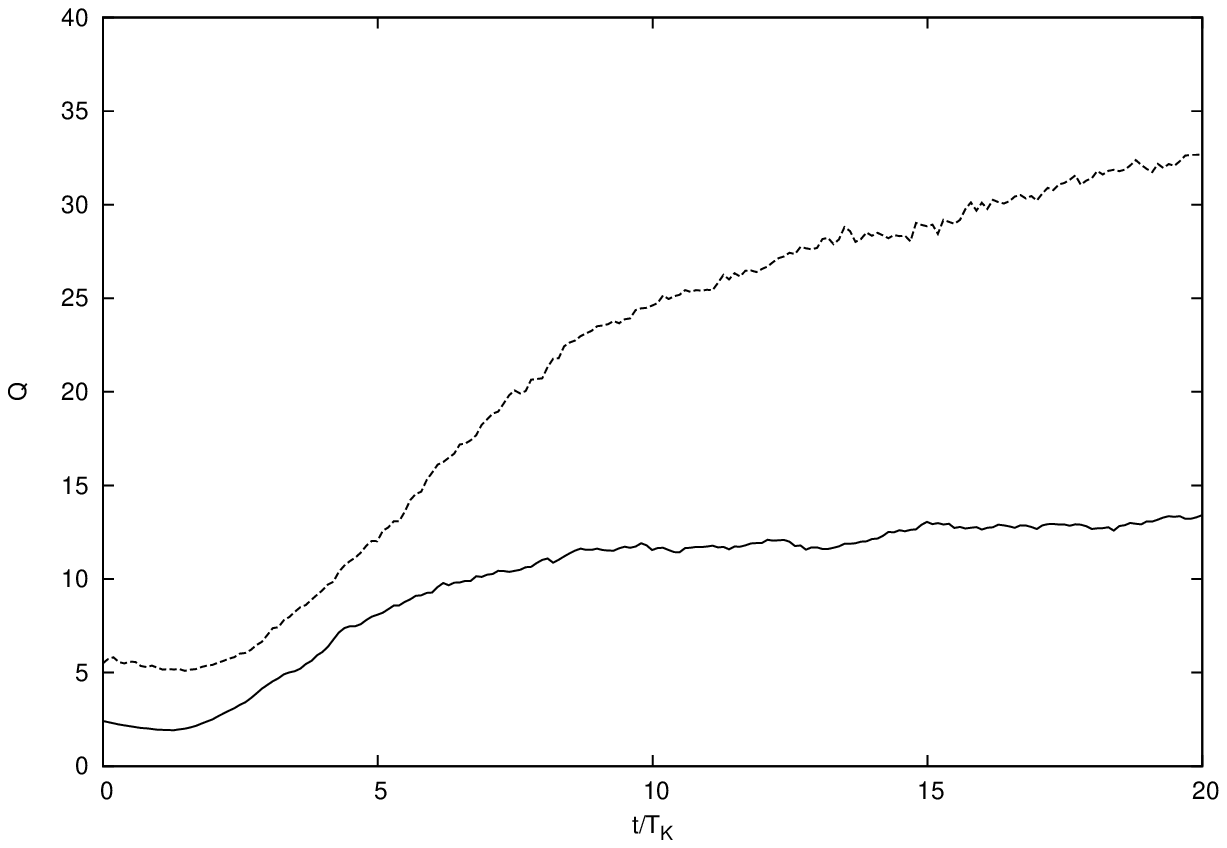} 
\end{center}
\caption{Time evolution of Toomre's $Q$ value $Q$ (solid line) and the ratio of the Roche density to the dust density $Q_R$ (dashed line) for model 1.  }
\label{fig:iniq.eps}
\end{figure}

The strength of self-gravity is measured by two dimensionless quantities: \citep[e.g.,][]{Youdin2005a},
\begin{equation}
  Q = \frac{\sigma_x \Omega_0}{\pi G \Sigma},
\end{equation}
\begin{equation}
  Q_R = \frac{ \rho_\mathrm{R}}{ \rho } \simeq \frac{9}{4\pi} \frac{h_d \Omega_0^2}{G \Sigma } ,
\end{equation}
where $\rho$ is the mass density of the dust layer, and $h_d$ is the thickness of the dust layer.  
The value $Q_R$ is the ratio of the Roche density to the dust density.
If these values are sufficiently small, the disk is
gravitationally unstable.  Figure \ref{fig:iniq.eps} shows time
evolution of $Q$ and $Q_R$.   
The value $Q_R$ is always larger than Toomre's $Q$ value.
The ratio  $Q / Q_R$ is approximately equal to  $0.5$.
If the hydrostatic balance ($\sigma_x  \simeq \Omega_0 h_d$) can be assumed, $Q / Q_R = 4/9 \simeq 0.44$.


If the gravitational instability occurs, there are no remarkable qualitative differences in the planetesimal formation process among different collision models.
The growth of particles in the accretion model is more efficient than in the hard and soft sphere models, thus 
the maximum mass of planetesimals is slightly larger than that in the hard and soft sphere models.


\subsection{Parameter Dependence}
\subsubsection{Physical Parameters}
We summarize the parameter dependences of the simulation results on 
the restitution coefficient $\epsilon$, the optical depth $\tau$, the ratio of
Hill radius to the particle diameter $\zeta$,
the initial Toomre's $Q$ value.  We focus on the mass of the largest particle, and Toomre's $Q$ value.

\begin{figure}
\begin{center}
\includegraphics[width=0.4\linewidth]{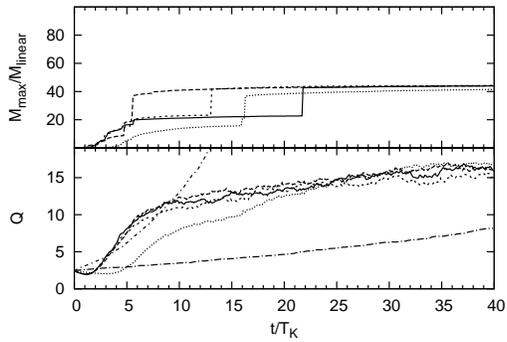} 
\end{center}
\caption{
Time evolution of various quantities for $\epsilon=0.01$ (\textit{solid curve}), $\epsilon=0.1$ (\textit{dashed curve}), $\epsilon=0.3$ (\textit{short dashed curve}) , $\epsilon=0.5$ (\textit{dotted curve}), $\epsilon=0.7$ (\textit{dash-dotted curve}), $\epsilon=0.9$ (\textit{dot-short-dashed curve}) (models 1,4,5,6,7,8).
The quantities are the mass of the largest body, and Toomre's $Q$ value from top to bottom, respectively.}
\label{fig:gi2dep_epsilon_all.eps}
\end{figure}

Figure \ref{fig:gi2dep_epsilon_all.eps} shows the dependence on the restitution coefficient $\epsilon$.  
We vary the restitution coefficient $\epsilon$ from $0.01$ to $0.9$ (models 1,4,5,6,7,8).  
No gravitational instability occurs, in other words, $Q$ never becomes less than about 2, for $\epsilon = 0.9$, and no large
body form.
There are two reasons why planetesimals do not form.
The velocity dispersion increases monotonically if $\epsilon$ is larger than the critical value $\epsilon_c \simeq 0.7$ \citep{Goldreich1982,Salo1995,Daisaka1999,Ohtsuki1999}. 
The velocity dispersion decreases due to inelastic collisions and increases due to gravitational scattering.
If the restitution coefficient is  large, the dissipation is relatively weak, thus the velocity dispersion does not decrease.
Because $Q$ value does not reach the critical value, gravitational instability
does not occur. Therefore, planetesimal seeds do not form, and the
subsequent collisional growth does not happen. We cannot apply the formation
process of planetesimal stated in our paper to the model with $\epsilon=0.9$. With a
larger restitution coefficient, the reduction of the relative velocity on
collisions is smaller. In addition, the collision velocity increases as the
velocity dispersion increases. The coagulation is not efficient with the large
restitution coefficient. Thus planetesimals do not form by coagulation within the
simulation time.   

\begin{figure}
\begin{center}
\includegraphics[width=0.4\linewidth]{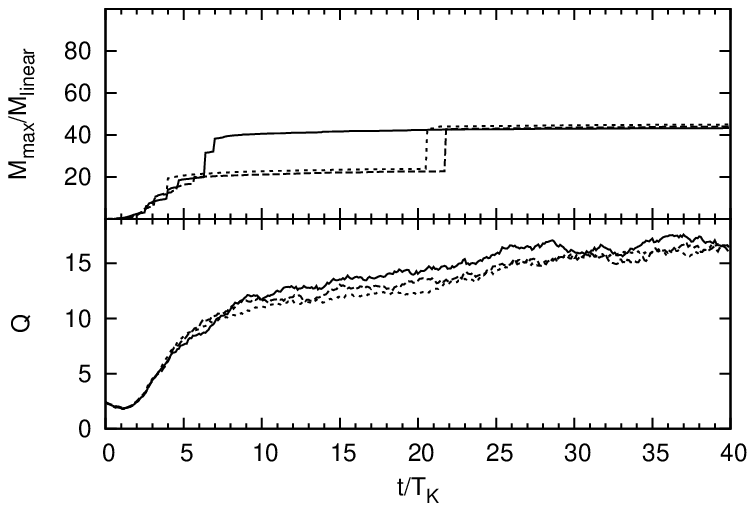} 
\end{center}
\caption{
Same as Figure \ref{fig:gi2dep_epsilon_all.eps} but for 
$\tau=0.1$ (\textit{solid curve}), $\tau=0.11$ (\textit{dashed curve}), $\tau=0.125$ (\textit{dotted curve}) (models 1,2,3).
}
\label{fig:gi2dep_tau_all.eps}
\end{figure}

We study the effect of the optical depth $\tau$ by comparing results with $\tau=0.1, 0.11,$ and $0.125$ (models 1,2,3).  
The dependence on the optical depth $\tau$ is shown in Figure \ref{fig:gi2dep_tau_all.eps}.  
The results are similar to those of the hard and soft sphere models (Paper I).  
The largest masses and Toomre's $Q$ value are on the same order.

\begin{figure}
\begin{center}
\includegraphics[width=0.4\linewidth]{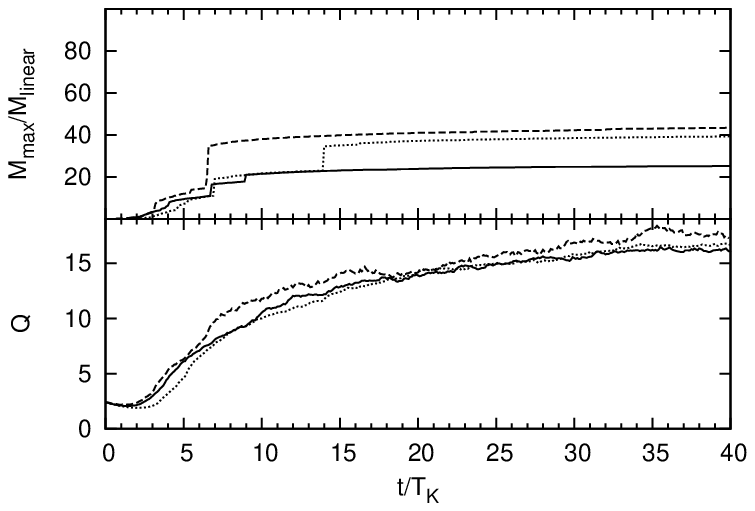} 
\end{center}
\caption{Same as Figure \ref{fig:gi2dep_epsilon_all.eps} but for 
$\zeta=2.5$ (\textit{solid curve}), $\zeta=2.75$ (\textit{dashed curve}), $\zeta=3.0$ (\textit{dotted curve}) (models 9,10,11).
}
\label{fig:gi2dep_hill_all.eps}
\end{figure}

We set $\zeta=2.5$, $2.75$, and $3.0$ and fix the other parameters in the standard model except for the optical depth $\tau=0.06$ 
(models 9,10,11). 
The dependence on $\zeta$ is shown in Figure \ref{fig:gi2dep_hill_all.eps} .
The result is also similar to that of the hard and soft sphere models (Paper I).
The largest masses are  $M_\mathrm{max}/M_\mathrm{linear}=43.3, 43.2,$ and $39.8$ respectively.
The largest masses are almost the same, which is about $40 M_{\mathrm{linear}}$.
Figure \ref{fig:gi2dep_hill_all.eps} shows the similar time evolution of
Toomre's $Q$ value because the masses of planetesimals are almost the same.

\begin{figure}
\begin{center}
\includegraphics{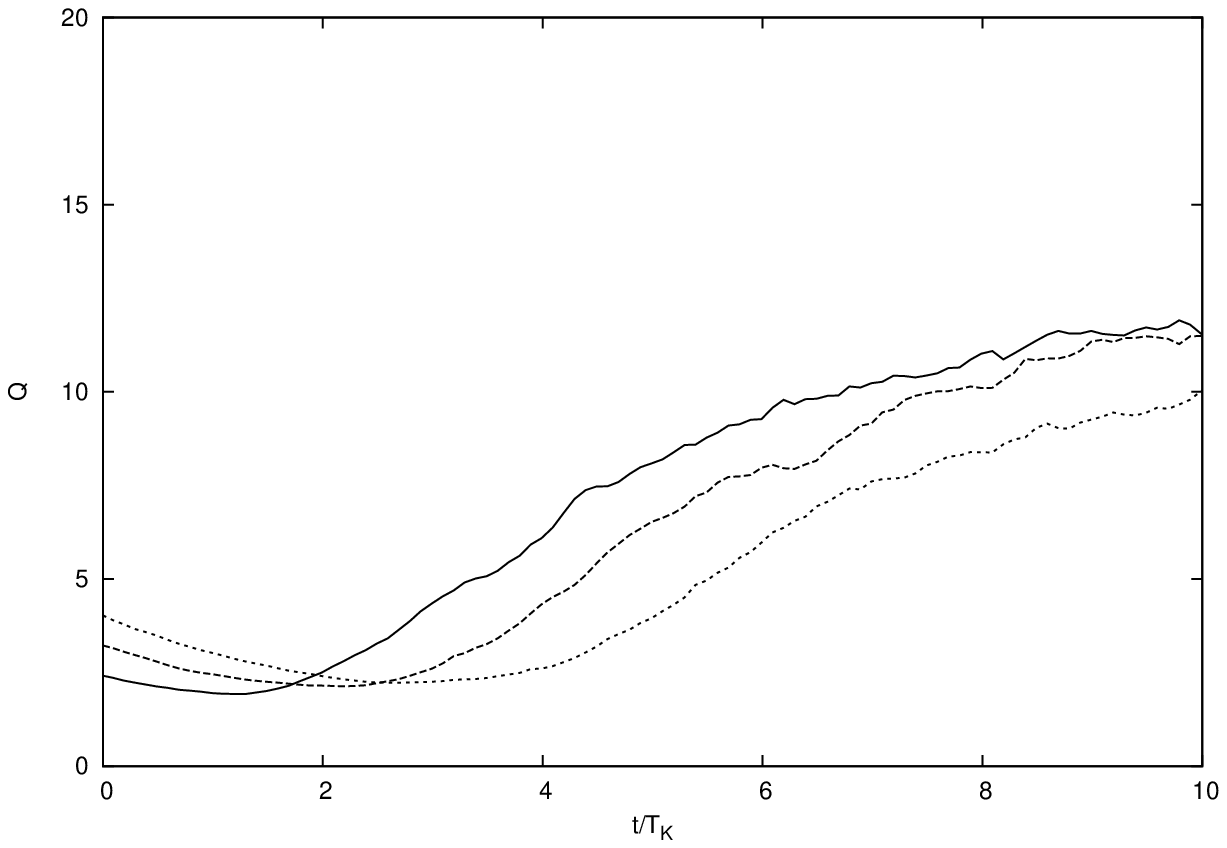} 
\end{center}
\caption{Time evolution of Toomre's $Q$ value. Initial $Q$ values are $3.0$ (\textit{solid curve}), $4.0$ (\textit{dashed curve}), $5.0$ (\textit{shot dashed curve}) (models 1, 23, 24).}
\label{fig:f51.eps}
\end{figure}

Figure \ref{fig:f51.eps} shows the time evolution of Toomre's $Q$ values for
$Q_\mathrm{init} = 3.0, 4.0,$ and $5.0$ (models 1, 23, 24).  If the growth is due to
gravitational instability, the results do not depend on the initial Toomre's
$Q$ value $Q_\mathrm{init}$.  Time evolutions are similar for these models.
Toomre's $Q$ value decreases through inelastic collisions between particles. As $Q$
decreases, self-gravity becomes stronger.  When $Q$ becomes less than about 2,
the disk becomes gravitationally unstable and the wake-like structure develops.
Then $Q$ increases due to scattering of field particles by planetesimal seeds.

In Paper I, we showed that there is no remarkable difference
between the hard and soft sphere models on the formation process of planetesimals.  
We confirm that the results of the accretion model are also essentially the same as these of the hard and soft sphere models.
The dissipation rate due to collision
is a crucial parameter but the planetesimal formation process is independent of
the collision model.  Before the gravitational instability, collisions merely
decrease the velocity dispersion.  In the late stage, coalescence among
planetesimal seeds occurs, but this process is independent of collision models.

\subsubsection{Simulation Domain Size}
In the hard and soft sphere models, 
the mass of the largest planetesimal depends on the size of the computational domain, and we could not find the saturation of growth (Paper I). 
The growth of planetesimals continues until most of the mass in the computational domain is absorbed by the largest planetesimal.
We expect to find the saturation of growth because the accretion model enables us to perform a wider simulation with a large number of particles.

\begin{figure}
\begin{center}
\includegraphics[width=0.4\linewidth]{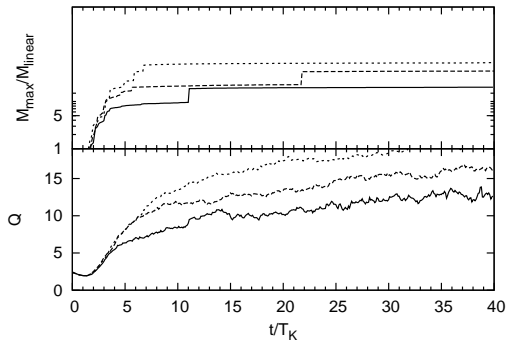} 
\end{center}
\caption{
Same as Figure \ref{fig:gi2dep_epsilon_all.eps} but for  $A_x=A_y=4$ (\textit{solid curve}), $A_x=A_y=6$ (\textit{dashed curve}), and $A_x=A_y=8$ (\textit{dash-dotted curve})(models 12,1,13)
.
}
\label{fig:gi2dep_region_all.eps}
\end{figure}

First we assume that the ratio of the length of the computational domain in the $x$-direction to the $y$-direction is unity, 
in other words, the shape of the computational domain is a square.
We vary the size of the computational domain $A=4$ to $8$.
Results are shown in Figure \ref{fig:gi2dep_region_all.eps} (models 12, 1, 13).
The final masses of the largest planetesimals and Toomre's $Q$ values are $M_\mathrm{max}/M_\mathrm{linear}=19.5, 44.0, 66.5$, $Q=13.6, 18.2, 21.9$, respectively. 
As the size of the simulation domain increases, the final planetesimal mass and the final Toomre's $Q$ value increase.

\begin{figure}
\begin{center}
\plottwo{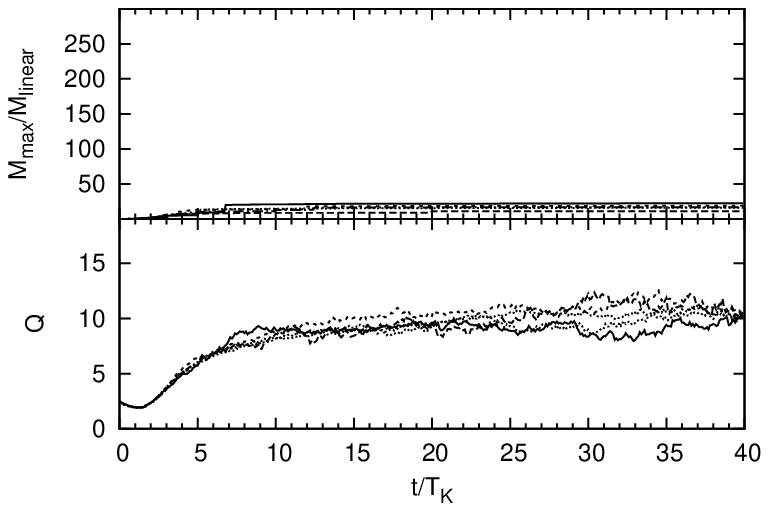}{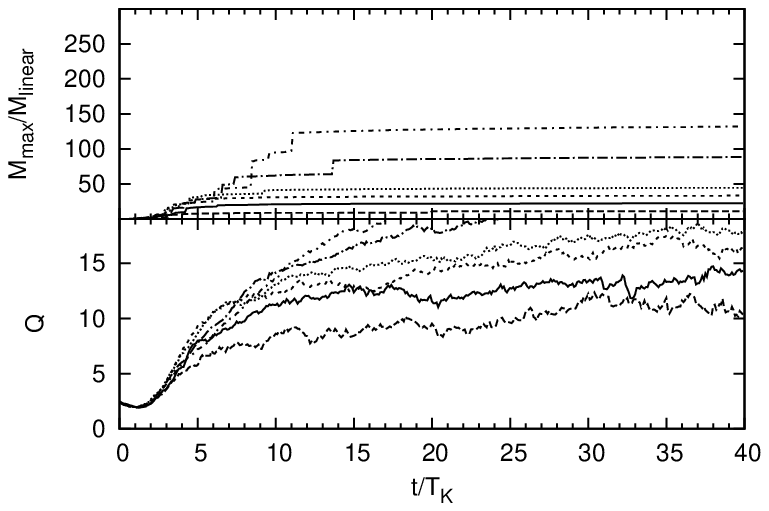} 
\end{center}
\caption{
Same as Figure \ref{fig:gi2dep_epsilon_all.eps} but for $A_x=3$ (\textit{solid curve}), $A_x=6$ (\textit{dashed curve}), 
$A_x=9$ (\textit{dash-dotted curve}), $A_x=12$ (\textit{short dashed curve}), 
and $A_x=24$ (\textit{dotted curve}) (models 14,15,16,17,18) in the \textit{left panel},
and $A_y=3$ (\textit{solid curve}), $A_y=6$ (\textit{dashed curve}), 
$A_y=9$ (\textit{dash-dotted curve}), $A_y=12$ (\textit{short dashed curve}), 
and $A_y=24$ (\textit{dotted curve}) (models 14,19,20,21,22) in the \textit{right panel}.
}
\label{fig:gi2dep_x_dir_all.eps}
\end{figure}

Next, we vary the ratio of the length of the computational domain in the $x$- and $y$-directions.
Now the shape of the computational domain is rectangular.
We vary the length in the $y$-direction, $A_y$, from 3 to 24 fixing the length in the $x$-direction, $A_x$, (the left panel in Figure \ref{fig:gi2dep_x_dir_all.eps}) (models 14, 15, 16, 17, 18), and
we vary $A_x$ from 3 to 24 fixing $A_y$ (the right panel in Figure \ref{fig:gi2dep_x_dir_all.eps}) (models 14, 19, 20, 21, 22).
The masses of the largest planetesimal and Toomre's $Q$ values are $M_\mathrm{max}/M_\mathrm{linear}=11.1, 21.7, 18.8, 15.9, 16.8$, $Q=10.4, 13.6, 10.2, 12.2, 11.3$ (models 14, 15, 16, 17, 18) 
and $M_\mathrm{max}/M_\mathrm{linear}=11.1, 21.8, 32.2, 43.1, 85.5$, $Q=10.4, 17.1, 15.9, 21.3, 24.1$ (models 14, 19, 20, 21, 22), respectively.
In the model with fixed $A_x$ , the mass of the largest planetesimal increases with $A_y$, while in the model with fixed $A_y$, the largest mass is roughly constant.  
These results indicate that the mass of the largest planetesimal is determined by $A_y$.
If $A_x$ is sufficiently large, multiple planetesimals form.
Toomre's $Q$ value increases with $A_y$ because of the scattering of field particles by the large bodies.

Here we estimate the final mass of a planetesimal,
$M_\mathrm{pl}$, in the computational domain in a similar way to estimate the isolation mass of protoplanets in the oligarchic growth picture \citep{Kokubo1998}. 
We assume that particles whose orbital radii are within $\gamma r_\mathrm{H}$ of the
planetesimal are finally absorbed by the planetesimal though they are not in the Hill sphere of the planetesimal initially, where $\gamma$ is a
non-dimensional factor and $r_\mathrm{H}$ is the Hill radius of the
planetesimal. Therefore, the planetesimal absorbs particles in the rectangle of $2 \gamma
r_{\mathrm{H}}$ in width and $L_y$ in length.
The hill radius of the planetesimal is $r_{\mathrm{H}} = a_0(2 M_\mathrm{pl}/3M_*)^{1/3}$, and we define 
\begin{equation}
  M_\mathrm{H}=2 \gamma r_{\mathrm{H}} L_y \Sigma.
  \label{eq:HillMass}
\end{equation}
The planetesimal with the mass $M_\mathrm{pl}$ can absorb the mass $M_\mathrm{H}$, 
and $M_\mathrm{H}$ increases as the planetesimal grows.
When $M_\mathrm{pl}\simeq M_\mathrm{H}$, the growth becomes slow. 
We estimate the mass of the planetesimal by this condition.
In this case the number of planetesimals is about $L_x/(2 \gamma r_\mathrm{H})$.
We obtained the mass of the planetesimal
\begin{equation}
  \tilde M_\mathrm{pl}=\min \left(576 \sqrt{6} \zeta^6 \tau^3 (\gamma A_y)^{3/2}, \tilde M_\mathrm{total} \right),
  \label{eq:planetesimal_mass}
\end{equation}
where $M_\mathrm{total}$ are the total mass in the computational domain. 
The number of the planetesimals is $N_\mathrm{pl}$ is
\begin{equation}
N_\mathrm{pl}= \max \left(\pi \sqrt{\frac{3}{8}} \frac{A_x }{\gamma^{3/2}A_y^{1/2}},1 \right).
  \label{eq:planetesimal_num}
\end{equation}
The mass of the planetesimals is equal to the total mass in the computational domain and the number of the planetesimal is unity 
if the following condition is satisfied:
\begin{equation}
  A_x < \frac{2 \sqrt{6}}{3 \pi} A_y^{1/2} \gamma^{3/2}.
  \label{eq:cond2}
\end{equation}

\begin{figure}
\begin{center}
\plottwo{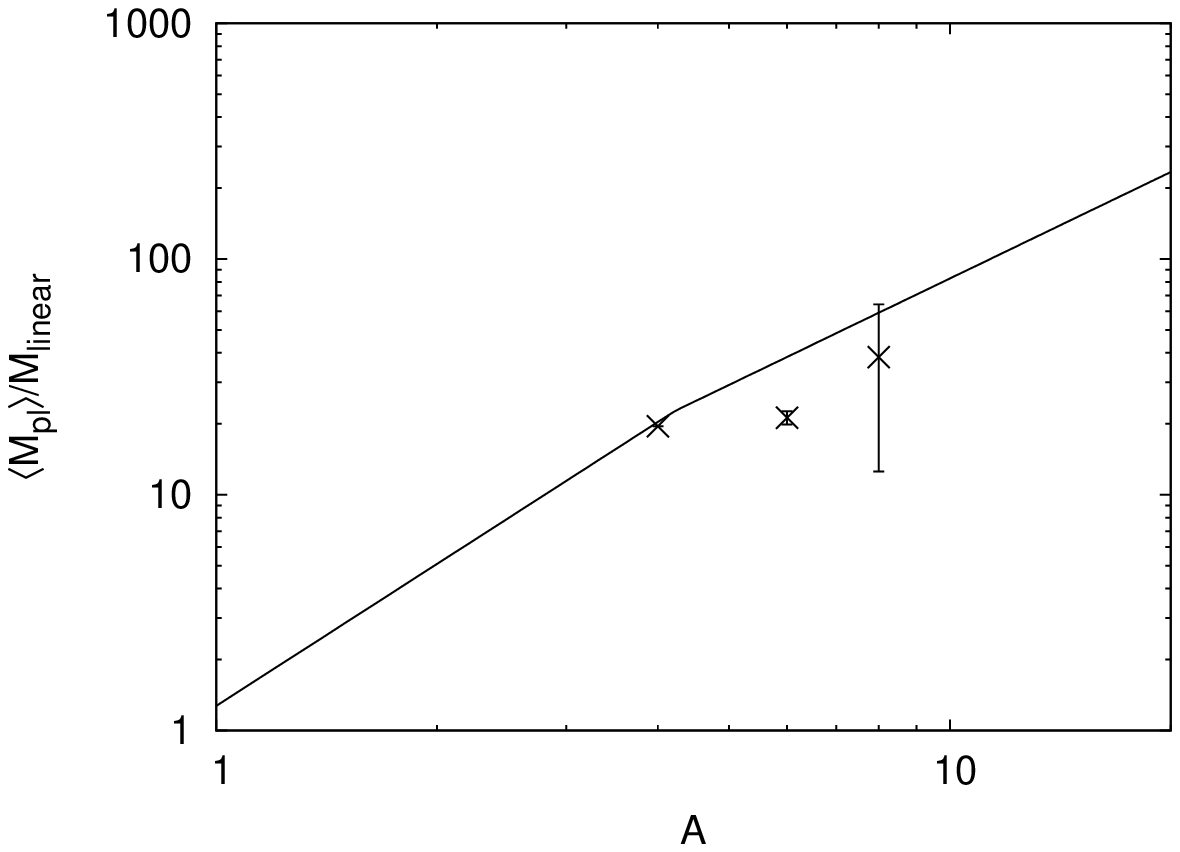}{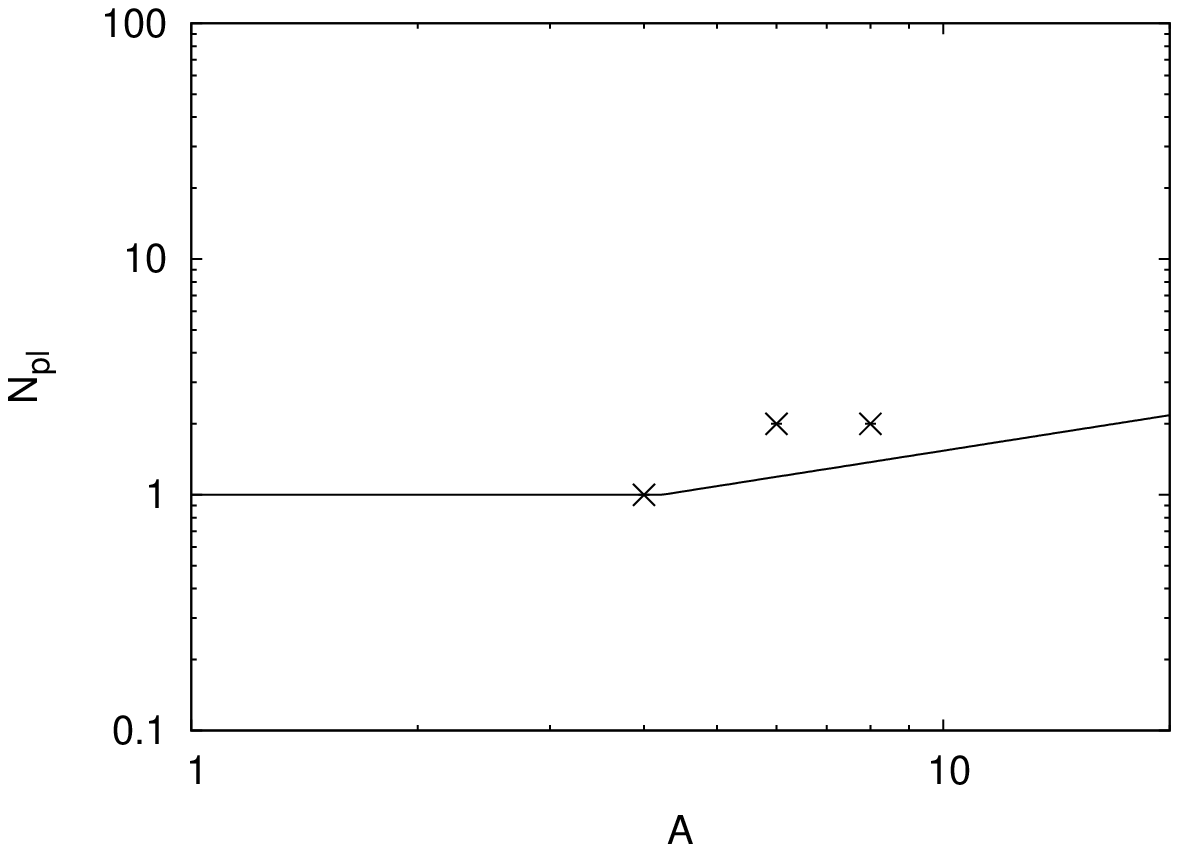}
\end{center}
\caption{Averaged mass $\langle M_{\mathrm{pl}} \rangle$ (\textit{left panel}) and the number $N_{\mathrm{pl}}$ (\textit{right panel}) of the planetesimals as a function of the size of the computational domain $A=A_x=A_y=4,6,8$ (models 12,1,13).  The solid line denotes the analytical estimation given by Equations (\ref{eq:planetesimal_mass}) and (\ref{eq:planetesimal_num}). We set $\gamma=2.5$.
The error bar shows the range of planetesimal masses in a given run.
}
\label{fig:plamax.eps}
\end{figure}

\begin{figure}
\begin{center}
\plottwo{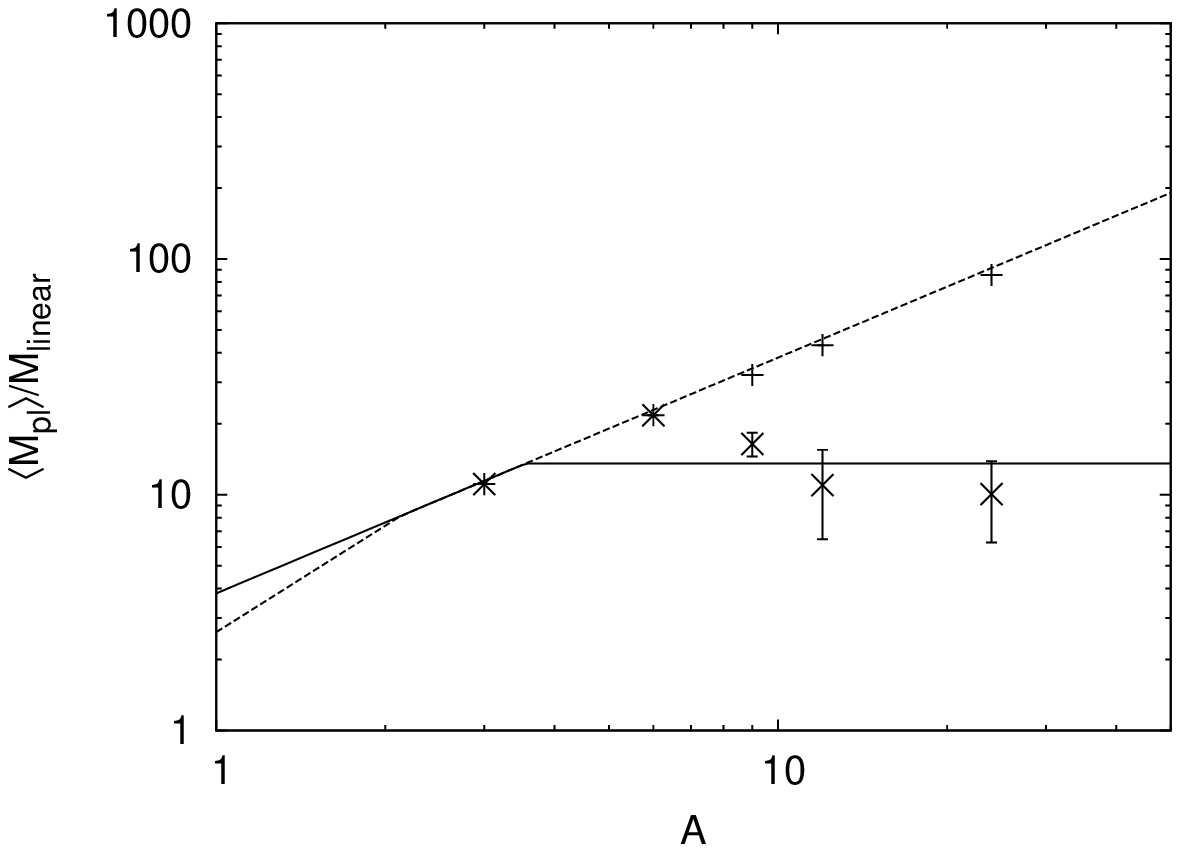}{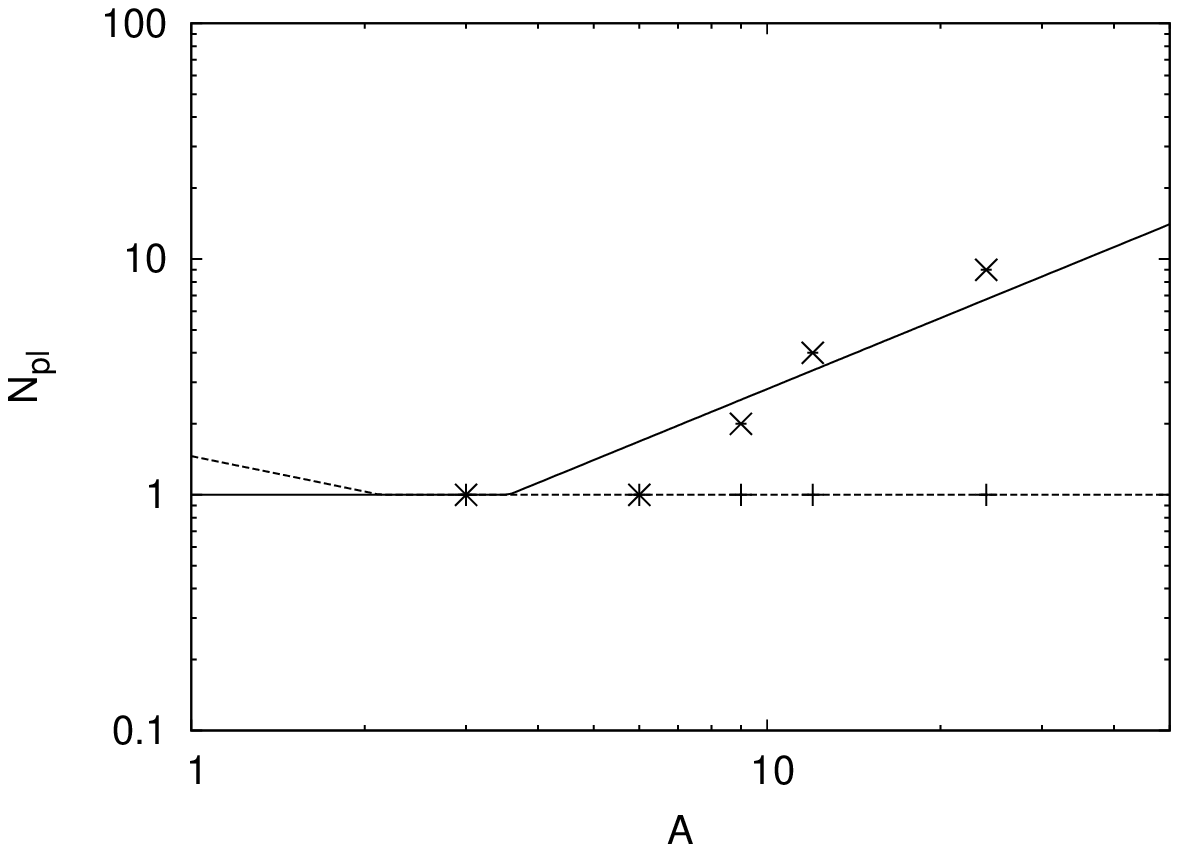}
\end{center}
\caption{Same as Figure \ref{fig:plamax.eps} but for $A=A_x=3,6,9,12,24$ (\textit{cross points}) (models 14,15,16,17,18) and $A=A_y=3,6,9,12,24$ (\textit{plus points}) (models 14,19,20,21,22). 
The dotted and solid lines denote the analytical estimations given by Equations (\ref{eq:planetesimal_mass}) and (\ref{eq:planetesimal_num}).
}
\label{fig:plamaxxy.eps}
\end{figure}

\begin{figure}
\plotone{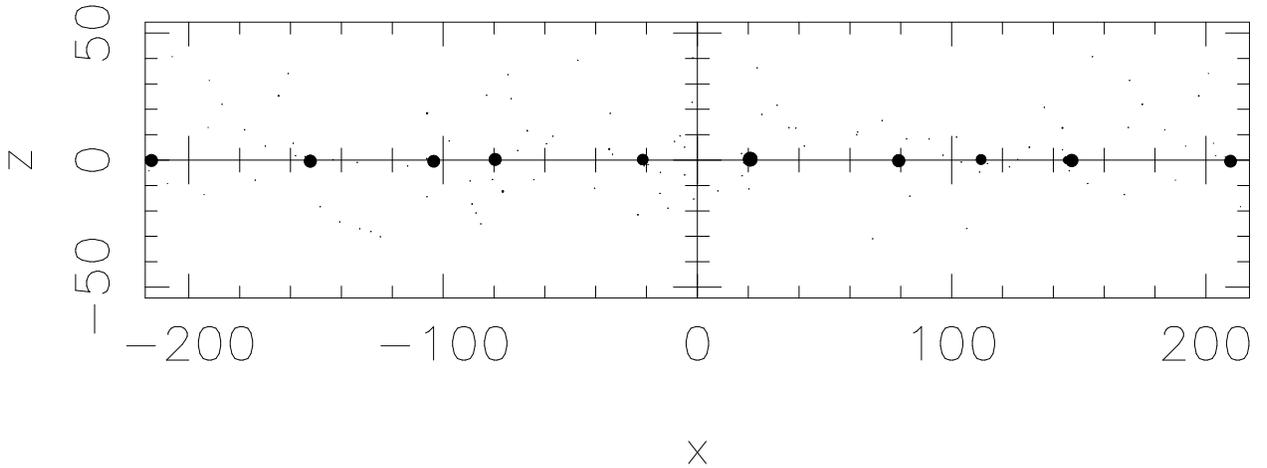} 
\caption{Spatial distribution of particles in model 18 at $t=40T_K$. The mean planetesimal mass is about $\langle M_\mathrm{pl}/M_\mathrm{linear} \rangle \simeq 10$ and its Hill radius is about $9.5$.  } 
\label{figz.eps}
\end{figure}

The comparisons between the analytical estimation and the simulations are shown in Figures \ref{fig:plamax.eps} and \ref{fig:plamaxxy.eps}. 
If we assume $\gamma=2.5$, the analytical estimation agrees with the simulation results.
The parameter $\gamma=2.5$ indicates that the mean orbital separation of planetesimals is $5 r_\mathrm{H}$. 
This is narrower than $10 r_\mathrm{H}$, which is the typical orbital separation of protoplanets caused by oligarchic growth \citep{Kokubo1998,Kokubo2000}.
This narrow orbital separation is possible for a dissipative system.
We will discuss this issue later.
For instance, Figure \ref{figz.eps} shows the final state in model 18. 
In this Figure, there are 9 planetesimals in the computational domain whose length in the $x$-direction is $400$.
Thus the mean orbital separation is about $44$.
The Hill radius of the mean planetesimal ($\langle M_\mathrm{pl}/M_\mathrm{linear} \rangle \simeq 10.1$) is about $9.5$.
Therefore the separation is estimated to about $9.5 \times 5 = 48$, which is consistent with Figure \ref{figz.eps}.

Our analytical estimation gives good agreement with the numerical results in each model.
Figure \ref{fig:plamax.eps} shows the mean mass and the number of planetesimals as a function of the size of the computational domain in the model where the domain is square. 
The mean mass of planetesimals increases with the size of the computational domain.
From the Equation (\ref{eq:cond2}), when $A_x=A_y<4.2$, the number of planetesimals is equal to unity and the mass
of the planetesimal is approximately equal to the total mass in the
computational domain. 
The slope of the line of the mean mass of planetesimals changes at $A_x=A_y=4.2$ and the number of the planetesimals is larger than unity if $A_x=A_y > 4.2$.

Figure \ref{fig:plamaxxy.eps} shows the result for the model of the rectangular computational domain.
In the models of $A_y=3$, the mean mass of planetesimals is constant and the number of planetesimals is larger than unity if $A_x>3.6$.
In the models of $A_x=3$, the slopes of the lines change at $A_y=2.1$.
The growth of planetesimals cannot be saturated although $A_y$ is large.
In these models, the Hill radius of the planetesimal is longer than the size of the computational domain in the $x$-direction.
The only one planetesimal forms, and it absorbs most mass in the computational domain.
It should be remembered that the local calculation of planetesimal formation has the domain-size dependence.

If we assume that the surface density of dust is 
\begin{equation}
  \Sigma = 10 \times f_\mathrm{d} \left( \frac{a_0}{1 \mathrm{AU}} \right)^{-3/2} \mathrm {g}\mathrm{cm}^{-2},
  \label{eq:hayashi}
\end{equation}
where $f_\mathrm{d}$ is a scaling factor, the linear mass $M_\mathrm{linear}$ is given by 
\begin{equation}
  M_\mathrm{linear}=3.5\times 10^{18} 
  f_\mathrm{d}^3
  \left(\frac{a_0}{1 \mathrm{AU}}\right)^{3/2}
  \left(\frac{M_*}{M_\odot}\right)^{-2}
  \mathrm{g},
\end{equation}
and from Equation (\ref{eq:planetesimal_mass}) the estimated planetesimal mass $M_\mathrm{pl}$ is given by
\begin{equation}
  M_\mathrm{pl}=9.0\times 10^{18} 
  f_\mathrm{d}^3
  \left(\frac{a_0}{1 \mathrm{AU}}\right)^{3/2}
  \left(\frac{M_*}{M_\odot}\right)^{-2}
  \left(\frac{\gamma}{2.5}\right)^{3/2}
  A_y^{3/2}
  \mathrm{g} = 2.6 \left(\frac{\gamma}{2.5}\right)^{3/2}
  A_y^{3/2}
  M_\mathrm{linear},
  \label{eq:planetesimalmass}
\end{equation}
where we assume $A_x > 2.1 A_y^{1/2} \left(\gamma/2.5\right)^{3/2}$.
The factor $f_\mathrm{d}=1$ roughly corresponds to the minimum mass solar nebula model inside the snow line \citep{Hayashi1981}.
From Equation (\ref{eq:planetesimalmass}), the estimated planetesimal mass becomes larger than the linear mass if the feeding zone is large, $(A_y \gamma)^{3/2} > 1.6 $.

The estimated and linear masses exhibit the same dependencies on $f_\mathrm{d}$, $a_0$, and $M_*$, and the estimated planetesimal mass is on the same order of the linear mass when $A_y \gamma \simeq 1$.  
These results are explained as follows.
Here we rewrite the Hill radius by the most unstable wavelength:
\begin{equation}
  r_\mathrm{H} = \frac{\lambda_\mathrm{most}}{(3 \pi)^{1/3}}  \left(\frac{M_\mathrm{pl}}{M_\mathrm{linear}} \right)^{1/3},
\end{equation}
where the most unstable wavelength for $Q=1$ is 
\begin{equation}
  \lambda_\mathrm{most} = \frac{2\pi^2 G \Sigma}{\Omega_0^2}.
\end{equation}
The most unstable wavelength of the gravitational instability and the Hill radius of the linear mass are on the same order.  
The condition $Q=1$ corresponds to the balance between the self-gravity and the effect of the rotation.
On the other hand, the Hill radius is determined by the balance between the self-gravity and the tidal force.
Thus, $r_\mathrm{H}$ and $\lambda_\mathrm{most}$ are on the same order, and the planetesimal mass is on the order of the linear mass only if $A_y \gamma \simeq 1$.
In general, because $A_y \gg 1$, the planetesimal mass is larger than the linear mass.

Collisions among planetesimals are more frequent than the realistic case in the present simulations since
the bulk density of super-particle is much smaller than the realistic value.
Therefore the orbital separation of large planetesimals is different from that of the standard oligarchic
growth derived by $N$-body simulations of planetesimals \citep{Kokubo1998,Kokubo2000}.  
If the number density is low, the orbital separation is determined by the balance between the orbital repulsion and their growth \citep{Kokubo1998}.
But if the number density is high, and the collisional dissipation is effective, the orbital repulsion is relatively weak. 
Thus, in the collisional oligarchic growth, the orbital separation can be narrower than $10R_H$ \citep{Goldreich2004}.

\section{Summary\label{sec:summary}}
We performed local $N$-body simulations of planetesimal formation through gravitational instability and collisional growth without a gas component
using a shearing box method.
The accretion model was adopted as the collision model.  
In the accretion model, when particles collide, if the binding condition is satisfied, two particles merge into one.
The number of particles decreases as the calculation proceeds, thus this model enables us to perform the long-term and large-scale calculations.
We compared the results of the accretion model with those of the hard and soft sphere models used in Paper I.

The formation process of planetesimals is the same as that of the hard and soft sphere models.
It is divided into three stages: the formation of non-axisymmetric structures, 
the creation of planetesimal seeds, and the collisional growth of the planetesimals.
The velocity dispersion of particles decreases gradually due to the inelastic collisions.
The gravitational instability occurs when the velocity dispersion of particles becomes a critical value determined by $Q \simeq 2$.
The non-axisymmetric structures form and fragment into planetesimal seeds, the masses of which are on the order of the linear mass. 
In this paper, we consider the formation of non-axisymmetric structures and planetesimal seeds  as gravitational instability.
Planetesimal seeds grow rapidly by coagulation and the large planetesimals form.
Most of the mass in the computational domain is absorbed by a few large planetesimals.

We studied the effect of physical parameters on planetesimal formation: the restitution coefficient $\epsilon$, the initial optical depth $\tau$, the dependence of planetesimal formation on the ratio $\zeta=r_\mathrm{H}/2r_\mathrm{p}$, and the initial Toomre's $Q$ value.
We found no qualitative differences among the collision models.
In the accretion model, the merged particles cannot split, thus the growth of particles is more efficient than in the hard and soft sphere models. 
If the restitution coefficient $\epsilon$ is smaller than the critical value, the time evolution is similar to the standard model.
If $\tau$ is large, the collision frequency is high, and the dissipation of the kinetic energy is efficient. 
Thus, the gravitational instability occurs more quickly for larger $\tau$.
In this narrow parameter range $\zeta = 2.5, 2.75, 3.0$, there is no remarkable dependence on $\zeta$.

By the long-term and large-scale calculations using the accretion model, we found that the mean mass of the planetesimals depends on the size of the computational domain.
We found that the mean mass of planetesimals is proportional to $L_y^{3/2}$. 
However, this mass is independent of $L_x$ if $L_x$ is sufficiently large.
The mean mass of planetesimals is estimated by Equation (\ref{eq:planetesimal_mass}).
Large planetesimals sweep small planetesimals in the rotational direction.
If the orbital separation of planetesimals is sufficiently large, the planetesimals cannot collide. 
The typical orbital separation is about $5 r_\mathrm{H}$ in the present simulation
The dependence of the planetesimal mass on the size of the computational domain indicates that we cannot simply discuss the realistic mass of the planetesimal using the local calculation.

The effect of gas was neglected in our simulation in order to study the
physical process of gravitational instability of dust particles and subsequent collisional evolution as a first
step.
Once gravitational instability happens and large aggregates form,  we may be able to neglect
gas drag because they are so large that they decouple from gas. 
So our gas-free model may be applicable to the stages of gravitational instability and
subsequent collisional growth.
The simple gas-free model provides thorough understanding of the
gravitational instability and collisional growth of planetesimals.  However, it is
necessary to include gas drag so as to investigate the realistic formation
process of planetesimals since the stopping time of small dust particles is
shorter than the Kepler time.  We will investigate the gravitational instability of
a particle disk embedded in a laminar gaseous disk in the next paper.

\acknowledgments{
We are thankful to an anonymous referee for helpful comments and suggestions.
This research was partially supported by MEXT (Ministry of Education, Culture,
Sports, Science and Technology), Japan, the Grant-in-Aid for Scientific
Research on Priority Areas, ``Development of Extra-Solar Planetary Science,''
and the Special Coordination Fund for Promoting Science and Technology,
``GRAPE-DR Project.''
S. I. is supported
by Grants-in-Aid (15740118, 16077202, and 18540238)
from MEXT of Japan.
}

\end{document}